\documentclass{article}       
\usepackage{lineno,hyperref}
\usepackage[english]{babel}
\usepackage[utf8]{inputenc}
\usepackage{graphics}
\usepackage{graphicx}
\usepackage{booktabs}
\usepackage{array}
\usepackage{paralist}
\usepackage{verbatim}
\usepackage{subfigure}
\usepackage{amssymb}
\usepackage{amsmath}
\usepackage{algorithm}
\usepackage{algpseudocode}
\usepackage{url}

\newcommand{\no}[1]{}
\newcommand{\NO}[1]{}
\newcommand{\set}[1]{\left\{#1\right\}}

\DeclareMathOperator{\knn}{\textsf{knn}}

\newcommand{\bftab}{\fontseries{b}\selectfont}
\begin{document}

\title{A scalable solution to the nearest neighbor search problem through local-search methods on neighbor graphs\thanks{A revised version of this manuscript has been published in Pattern Analysis and Applications, 24, 763–777 (2021). \url{https://doi.org/10.1007/s10044-020-00946-w}}}

\author{Eric S. Tellez$^1$\\ CONACyT-INFOTEC \\ \url{eric.tellez@infotec.mx} \and
Guillermo Ruiz$^2$\\ CONACyT-CentroGEO \\ \url{lgruiz@centrogeo.edu.mx} \and
Edgar Chavez$^3$ \\ CICESE \\ \url{elchavez@cicese.mx}
\and Mario Graff$^1$ \\ CONACyT-INFOTEC \\ \url{mario.graff@infotec.mx}}

\date{\scriptsize
$^1$~CONACyT-INFOTEC Centro de Investigación e Innovación en Tecnologías de la Información y Comunicación. Circuito Tecnopolo Sur 112, Fracc. Tecnopolo Pocitos, CP 20313. Aguascalientes, Ags, México\\
$^2$~CONACyT-CentroGEO Centro de Investigación en Geografía y Geomática
  ``Ing. Jorge L. Tamayo'', A.C. Circuito Tecnopolo Norte No. 117, Col. Tecnopolo Pocitos II, C.P. 20313. Aguascalientes, Ags, México.
  \\
  $^3$~CICESE Centro de Investigación Científica y de Educación Superior de Ensenada, Ensenada, Baja California, México
  }

\maketitle

\begin{abstract}
Near neighbor search (NNS) is a powerful abstraction for data access; however, data indexing is troublesome even for approximate indexes. For intrinsically high-dimensional data, high-quality fast searches demand either indexes with impractically large memory usage or preprocessing time.

In this paper, we introduce an algorithm to solve a nearest-neighbor query $q$ by minimizing a kernel function defined by the distance from $q$ to each object in the database. The minimization is performed using metaheuristics to solve the problem rapidly; even when some methods in the literature use this strategy behind the scenes, our approach is the first one using it explicitly. We also provide two approaches to select edges in the graph's construction stage that limit memory footprint and reduce the number of free parameters simultaneously.

We carry out a thorough experimental comparison with state-of-the-art indexes through synthetic and real-world datasets; we found out that our contributions achieve competitive performances regarding speed, accuracy, and memory in almost any of our benchmarks.

\end{abstract}



\section{Introduction}
\label{section/intro}
The {\em metric space} abstraction for {\em proximity search} can be applied to a large number of problems; for instance, it is a fundamental tool to solve problems from statistics, data mining and pattern recognition to multimedia information retrieval, machine learning, compression, biometric identification, and bioinformatics \cite{Ssisap2010}. The problem involves preprocessing a database of objects equipped with a distance to find the nearest neighbors of a query, or objects within a certain distance threshold. More formally, let $U$ be a domain and $S=\{s_1,\dots,s_n\} \subset U$ a finite subset of $U$, and let $d\colon U \times U \rightarrow \mathbb{R}^+$ a distance. 
Given a query $q \in U$, the similarity search problem consists of preprocessing $S$ to find objects close to $q$ quickly. The preprocessing produces a search data structure called index that has the objective of speeding up the computation of similarity queries over $S$. This manuscript focuses on solving $k$ nearest neighbor queries, written as $\knn(q)$, which retrieves the $k$ most similar items in $S$ to the query object $q$. We understand similar as {\em close} regarding the distance $d$.


Finding the $k$ nearest neighbors for a particular application depends of the metric function used. Different problems require different distance functions and it is not clear what will be more suitable in a particular scenario. An alternative is to train a custom metric using Metric Learning. Metric Learning is the process to learn a metric such that the objects of the same class have small distance and objects of different classes end far apart. This can be used on image retrieval, person re-identification, and classification in general.
Some methods are Large Margin Nearest Neighbor \cite{10.5555/1577069.1577078}, Neighborhood Component Analysis \cite{10.1145/1273496.1273523}, and recently Li et al \cite{7243340}, used deep learning over images labeled by the community to present a new algorithm, named weakly-supervised deep metric learning which is able to preserve the distance of visual features as well as the text labels.


Designing algorithms and data structures that efficiently solve the search problem for small intrinsic dimensions is possible, but there is a compelling body of theoretical and empirical evidence about the raising of the complexity for high dimensional data, the so called {\em curse of dimensionality} (CoD). This phenomenon arises in the distribution of the pairwise distances between objects; roughly, the mass concentrates around the mean value, making it difficult to distinguish objects using the distance function. The net effect is that, even for highly selective queries, it is not possible to build a data structure that avoids the evaluation of a large part of the database to solve a query. Another effect is that the mean value increases with the dimension, this is valid for distance functions computed accumulatively like most Minkowsky's distance functions.

As Chavez et al.~\cite{CNBYMacmcs2001} describe, most of the search structures found in the literature will suffer from the CoD dimensional effect. The primary strategy to cope with the CoD is to allow the retrieval of an approximate answer; the correctness of the approximation can be measure by an approximation factor. Another standard approximation scheme is to measure the number of errors found in a result set; in this scenario, an error implies that the solution proposed may omit objects of the correct response, or add some objects that are not part of the exact result set. These methods are prone to introduce {\em false positives} and {\em false negatives} errors. We are interested in the latter kind of approximation algorithms since the approximation factor can be tricked when the underlying distribution of pairwise distance functions have high mean values and high-concentrations of mass around the mean; this is worth to consider since these are the symptoms of the CoD. 

In this paper, we introduce a graph based algorithm to approximate a nearest-neighbor query $q$ by minimizing a kernel function defined by the distance from $q$ to each object in the database.

\subsection{Overview}
The next section puts in context our contribution giving a brief review of the related state of the art literature. The rest of the manuscript is organized as follows. Section~\ref{sec/related-work} is dedicated to review the related literature and contrast it with our approach. Section~\ref{sec/contribution} discusses the fundamental assumptions behind our contribution, the general ideas behind the adapted metaheuristics, our similarity search algorithm, and also our neighborhood selection approaches. In Section \ref{section/results}, we present experiments over synthetic and real-world databases, in particular, we provide a broad comparison under different benchmarking conditions like databases with increasing dimension and the effect of the database's size.
Finally, Section~\ref{sec/conclusions} concludes the manuscript and suggests future directions of research.

\section{Related work}
\label{sec/related-work}

{\em Locality Sensitive Hashing} (LSH) is among the most well-known techniques for approximate similarity search. It was first presented in~\cite{GIMvldb1999}, with many follow up work as in \cite{AIacmcomm2008,heo2015spherical,andoni2015practical}. An LSH scheme is composed of a set of specialized hashing functions that work for a particular problem, e.g., vectors under the Euclidean distance, cosine similarity, or sets under the Jaccard distance. The central idea of a hashing function is that close objects are mapped to the same hash value (they will be assigned to the same bucket) with high probability while distant objects are mapped to different hashes with high probability.
As usual, in this kind of problem, the complexity of LSH depends both on the dataset and the set of queries, such that a complex problem will need many hash functions to ensure the expected result's quality. That is because the hash functions are very often correlated, and their discriminate power is sub-linear.

Most of the LSH methods use projections to map the original space. In \cite{6296665} was presented ITQ, an iterative method to find useful binary codes for the points in the dataset. The codes can be seen as a vertex of the hypercube $\set{-1, 1}^c$ where $c$ is the length of the codes. The code associated with every point is the closest vertex on the cube. The iterations find a good transformation of the hypercube to approximate the data.
Recently, on \cite{ijcai2018-292} was shown the Complementary Binary Quantization (CBQ) method to learn hash functions jointly based on prototypes for multiple hash tables. Such prototypes are well distributed among the data; the hashed space's distance notion resembles that of the original space. That is possible because only one prototype is needed to find the nearest neighbor. So, the hash functions can be seen as a map from the original space to the Hamming space. One difference with the ITQ is that the hypercube's vertexes are more evenly distributed over the data, and therefore, the hash functions are less redundant. Also, on the subsequent paper \cite{8999816}, the authors show how this can be done in a distributed configuration producing the D-CBQ method for use on big datasets.


The {\em Fast Library for Approximate Nearest Neighbors}, FLANN \cite{FLANN}, is a library that implements several indexing strategies and selects the more suitable one, based on parameters like the search time, the construction time, and the available memory.
With those parameters, together with the parameters of the indexes, a cost function is defined. FLANN builds and tries different instances of the indexes using optimization techniques to find a local minimum to the cost function. The instance corresponding to the local minimum will be the one selected.

Product quantization (PQ)~\cite{jegou2011} relies on vector quantization of subspaces. The central idea is to apply a clustering algorithm, like k-means, in several independent subspaces. The clustering is applied to a sample of the dataset, encoding the rest of the database according to the k-means centers. The authors compare their approach with FLANN using the SIFT-1M benchmark; PQ is between two and five times faster than FLANN for the same quality of results. Centers serve as symbols or words, and then efficient indexing techniques like Inverted Indexes can be used to support large datasets.
In \cite{OptimizedProductQuantization2014} the encoding error of PQ is reduced through an iterative quantization that encodes the database as a binary Hamming space, in essence. The vector quantization idea is applied recursively in \cite{FastNN} improving PQ and the optimized PQ. Following this line, on \cite{liu2015structure} the authors proposed a hashing structure based on PQ to capture the local and global properties of the data. 

Another scheme is based on representing each object with an ordered set of its $k$ nearest references, where a reference is just an special type of object. For instance, we can recognize this approach on CNAPP \cite{TCNis2012}, PP-Index \cite{Eppindex2014}, MIF \cite{AGSmetric2014}, and the quantized permutations~\cite{MOHAMED2015}. The similarity among the sets that represent any pair of items hints the similarity of the actual objects.
Queries are resolved using an inverted index over the sets. This structural similarity was systematically explored in \cite{KNR2015}, adding several new indexes to the list. 
The role played by the selection of the set of references is detailed in \cite{AMATO2015}.

An alternative to palliate the CoD consists of the usage of a combinatorial solution. This is the kind of work presented in \cite{lifshits} where the authors create a navigation graph, and iteratively improve the distance from the current node to the query, moving to a closer node. 
These algorithms have theoretical interest, but its practical applications are limited due to their quadratic memory requirements.

An alternative combinatorial algorithm that reduces this memory issue is the {\em Rank Cover Trees} (RCT) \cite{RCT}. Instead of using a random object to start the search, the authors use a rooted tree. Node descendants in the tree are obtained using a rank order, and since only rank information is used for navigation, and the degree of the tree is bounded, the total number of nodes visited in each search can be fixed beforehand. 
The Spatial Approximation Sample Hierarchy (SASH) \cite{sash} introduces a multilevel data structure where each level is a random sample of half the size of the previous level. Each object in an upper-level connects to a number of its approximate nearest neighbors at a lower level. Queries are resolved top-down searching for the nearest neighbors. At each level, the degree of the tree is bounded, and the search is done top-down seeking the most promising nodes. This process makes the whole searching procedure bounded time-wise, but without proximity guarantees. 

\subsection{The navigable small world graph}
We detail the base of our contribution, Navigable Small World (NSW), introduced in \cite{MPLK12}. Instead of building a tree-like SASH, the authors propose a search graph incrementally built with consecutive insertions.
Note that each insertion is defined regarding the search algorithm with a simple rule: To insert the $j$-th element, find the (approximate) $N$ nearest neighbors among the $j-1$ elements already indexed; then, the new item is linked (in both directions) to its $N$ nearest neighbors. The authors encourage the preservation of links to distant objects since they can boost search performance. Notice that these {\em long-links} will naturally arise using incremental construction.

The search algorithm starts selecting a random point and greedily follows the neighbor that minimizes the distance to the query, repeating until no further improvement is possible. This simple procedure performs fast but poorly regarding the recall. However, the quality can improve with amplification, with $m$ independent searches from random starting points. In a follow-up paper \cite{MPLK14}, the searching procedure is updated to use persistent sets with the restarts. In other words, it stores the set of visited candidates along the entire search process. At the beginning of each restart, the algorithm appends a random item to the list of candidates to add diversity to the search process.

The total number of evaluated distance functions of the search is determined by $m \cdot N \cdot {hops}$; where $m$ is the number of restarts, $N$ the degree of each node in the graph, and {\em hops} is the average number of hops needed to stop the search procedure.  In particular, the $m$ and $N$ parameters control both accuracy and speed and need to be established for each dataset.
While large $m$ values can improve the result's quality, too many restarts may decrease the overall search speed. Moreover, increasing $N$ could also boost recall and speed at the expense of using more memory, which can become an issue whenever we are dealing with large datasets. Also, the search speed is not linear in $N$, since other parameters become affected in the full expression, and surpassing the best value of $N$ may also lead to bad performances. The original paper experimentally shows, for their benchmarking datasets, that fixing $m=O(\log n)$ and $N=O(\log n)$ for $n$ vertices it produces $O(\log^3{n})$ searches, i.e., the number of hops is also $O(\log n)$.

Since NSW is built incrementally using object insertions, and each addition is defined regarding the search algorithm; therefore, there is an unusual feedback loop involved. A better approximation to the actual nearest neighbors may produce a different underlying graph, and with different characteristics.

In \cite{ruiz2013local} we develop a beam search based algorithm, that improves the performance of NSW in many cases; however, in addition to the beam size parameter and the neighborhood size, it needs to setup a small value $\sigma$ that is used to automatically stop the search.

A significant improvement to the search graph is made in \cite{malkov2018efficient}, where a Hierarchical Navigable Small World (HNSW) is introduced. The core idea is to construct a hierarchy of navigable nets connecting the nearest objects between layers. Each layer is denser than the previous one. A search is then solved following nearest neighbors per layer, starting in the less dense one, and following the connections to improve the approximation procedure. In addition to the NSW, the HNSW requires taking care of the list of layers and its connections. For neighborhood selection, HNSW uses a heuristic based on the Spatial Access Tree (SAT, see \cite{SAT}). The SAT-based heuristic selects a sub-list of a large list of neighbors. It starts with the node $u$ being inserted and a list of candidate neighbors $L$; $L$ is a list of {\em efConstruction} nearest neighbors of $u$ computed using the index, $L$ is ordered by distance to $u$.  The actual neighborhood $N$ is computed, iteratively, as follows:
\begin{itemize}
    \item For each $p$ in $L$ in order, compute the nearest neighbor to $p$ among items in the neighbor $N$ and also $u$.
    \item If $u$ is the nearest neighbor of $p$ then $p$ is added into the neighborhood $N$.
\end{itemize}

An HNSW user needs to tune two parameters, the {\em efConstruction} and $M$, the maximum size of the neighborhood (truncated for neighborhoods larger than $M$).

\subsection{Our contribution}

We devise that the search graph, as mentioned above, is a combinatorial search-space by itself. Then, the core idea is to navigate the graph using local information and the accumulated knowledge of previous evaluations until a local minimum is reached. We use several strategies to avoid local minimum solutions, and these approaches can improve the quality of the result set significantly.

Our approach is to adapt local-search methods, borrowed from combinatorial optimization, to solve nearest neighbor queries. This paper also studies heuristics for reducing parameter tuning, and in particular, on neighborhood selection heuristics. We characterize the search speed and quality of our search structures based on the performance of synthetic and real-world datasets. In the road, we also provide an extensive experimental comparison of our indexes with state of the art methods.

\section{Local-search methods to similarity search}
\label{sec/contribution}
\begin{figure}[!th]
\centering
\includegraphics[width=0.6\textwidth]{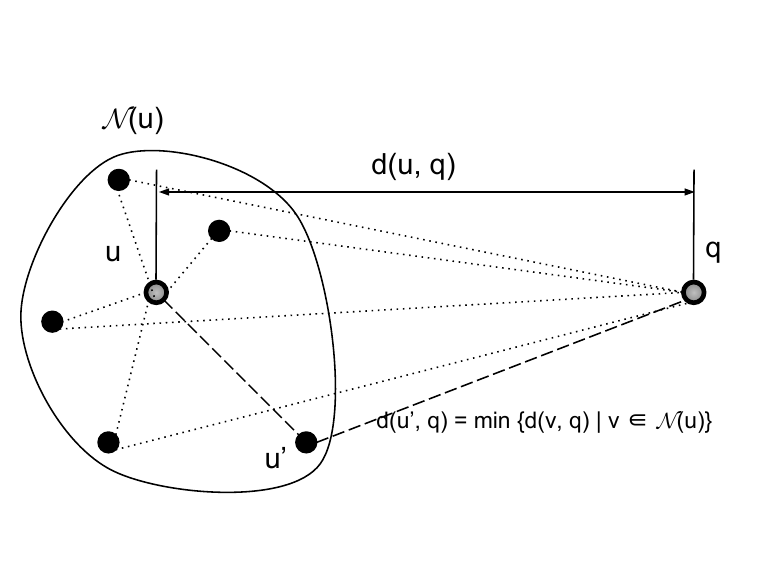}
\caption{An object $u$ and its neighborhood $\mathcal{N}(u)$;  $u'$ is the item closer to $q$ in $\mathcal{N}(u)$.}
\label{fig/evalneighbors}
\end{figure}

Our approach explores the improvement of the search algorithm of NSW based in that the search graph defines a search space for an underlying combinatorial optimization problem. We claim that each query defines an optimization landscape, using distance among the query to each object in the dataset. That way, the process of finding a minimum value solves (approximately) its associated nearest neighbor query. 
Figure \ref{fig/evalneighbors} illustrates the central assumption of the search graph to solve a query; the idea is that a query $q$ can be solved navigating the graph starting at some vertex $u$, such that our best initial guess of $\knn(q)$ is $u$. We can improve our current estimate exploring the neighborhood of $u$, $\mathcal{N}(u)$, selecting the best result, $u'$ in the figure. This process is repeated until there is no better object that minimizes the smaller known distance. Since we can be trapped in a local minimum, we need to perform different strategies to find better approximations. A priority queue of fixed size is required to store the most promising results to answer the $\knn(u)$ query. Using variations of this general scheme, we can create different search graphs, as will be detailed below.

Before we detail our contribution, we dedicate the next subsection to review the generic metaheuristics that were used in our approach.

\subsection{Local search heuristics}
\label{sec/heuristics}

As commented, one of the main issues in the scheme is to get away from local minima; that is the reason why several heuristics exist. There is a plethora of {\em local search} heuristics and metaheuristics besides the greedy search, here we focus on those used in our approach. The interested reader is referenced to the excellent book \cite{BurkeKendall2014} on the field. 

{\em Greedy search.}
It is interesting to revisit the Greedy Search (GS) strategy, identified as the heuristic behind NSW. If each try in the greedy search has a probability $P$ of success, it can be amplified to any target success probability of $P^\star$ using $m$ independent tries. This amplification is a standard method in combinatorial optimization, and it is proved that $m = O({\log{(1-P^\star)}}/{\log {(1 - P)}})$. For the problem of near-neighbor search, success probability $P$ can be assumed to be the recall, and be measured experimentally. It is interesting to notice that, under this assumption, any positive recall can be amplified with enough tries. The limitation, however, is the assumption of independence; so, if the greedy searches are not independent, the number of tries can be much larger to achieve the desired recall. The amplification process is also known as Iterated Hill Climbing (IHC).

{\em Tabu search} is a metaheuristic to avoid greedy decisions at each step. This strategy is designed to escape from local minima and plateaus. The central idea is to avoid taking the same path more than once; this is accomplished by marking already visited vertices as {\em prohibited} and avoid visiting them on later stages of the search process.

{\em Beam Search} metaheuristic stores a beam $B$ of promising candidates (already visited in the search process) of fixed size $b$. At each step, the beam search computes a new beam $B' = \cup_{s\in B}\mathcal{N}(s)$ and selects the closer objects from $B$ and $B'$ (the new beam replaces the old one). The search stops whenever the current beam $B$ cannot be improved. Tabu search can be used in conjunction with beam search to avoid non-essential computations. At the beginning of the search process, $B$ is populated with random vertices from the graph.

\subsection{Our search algorithms}
\begin{algorithm}
\begin{algorithmic}[1]
\Require The arguments of the function are: distance function $\textsf{dist}$, the search graph $G=(S,\mathcal{N}(S)$, the query $q \in U$, the number of neighbors $k$, and the beam size $b$.
\Require The algorithm needs the following functionality:
\begin{itemize}
    \item $R$ and $B$ are priority min-queues of fixed size that store pairs $(\textsf{dist}(q,u),u)$, prioritized by distance to the query,
    \item the $\textsf{nearest}$ function retrieves the pair with the minimum associated item, an analogous $\textsf{farthest}$ function,
    \item the $\textsf{popnearest}$ function that removes the closer pair in the queue,
    \item pairs support comparison by distance, between pairs and raw distance values,
    \item a pair $p$ can be inserted into the queue $Q$ when the queue has not reached its maximum size or when $p < \textsf{farthest}(Q)$.
\end{itemize}
\Ensure The result set $R$ containing the $k$ nearest neighbors of $q$.
\Function{BeamSearch}{$\textsf{dist}, S, \mathcal{N}, q, k, b$}
\State initialize $R$ as a priority min-queue of maximum size $k$.
\State initialize $B$ as a priority min-queue of maximum size $b$.
\State $states \leftarrow \emptyset$ \Comment{a dictionary to mark vertices as \textsf{visited} or \textsf{explored}}

\For{$i=1$ to $b$}  \Comment{initializes the search with $b$ random points}
    \State randomly select $u$ from $S$
    \If{$u \not\in states$} \Comment{avoids already visited points}
        \State $states_u \leftarrow \textsf{visited}$ \Comment{marks $u$ as visited}
        \State insert $(\textsf{dist}(u, q), u)$ into $R$ \Comment{pushes $u$ into the result set}
    \EndIf
\EndFor

\Repeat \Comment{navigate the graph while the result improves}
    \State $prev \leftarrow \textsf{farthest}(R)$
    \State insert $\textsf{nearest}(R)$ into $B$ \Comment{seeds $B$ with the current best neighbor}
    \While{$|B| > 0$} \Comment{explore the beam}
        \State $p \leftarrow \textsf{popnearest}(B)$ \Comment{removes best approximation in $B$ to explore it}
        \If{$state_p \not= \textsf{explored}$} \Comment{ignores already explored vertices}
            \State $state_p \leftarrow \textsf{explored}$ \Comment{marks $p$ as explored}
            \For{$c \in \mathcal{N}(p)$} \Comment{iterate each child $c$ in the neighborhood of $p$}
                \If{$c \not\in states$} \Comment{avoids already visited elements}
                    \State $states_{c} \leftarrow \textsf{visited}$ \Comment{marks $c$ as visited}
                    \If{$(\mathsf{dist}(q, c), c)$ can be inserted into $R$}
                        \State insert $(\mathsf{dist}(q, c), c)$ into $R$ and $B$
                    \EndIf
                \EndIf
            \EndFor
        \EndIf
    \EndWhile
\Until{$prev \geq \textsf{farthest}(R)$} \Comment{stops when result set does not improve}

\State \Return $R$
\EndFunction
\end{algorithmic}
\caption{Similarity search with our beam search algorithm.}
\label{alg/bs}
\end{algorithm}

Our search algorithm is Beam Search (BS) (see Algorithm~\ref{alg/bs}). The idea behind BS is to populate a priority queue of fixed-size $B$, called {\em beam}, that contains possible results and refines them iteratively.
Note that our algorithm uses a kind of memory to remember visited nodes like tabu search does (line 4). Lines 5-11 initialize the search procedure.  The exploration of the beam (lines 15-28) evaluates the neighborhoods of all objects in the beam, and it is replaced using those objects with a smaller distance to the query object. When the exploration finishes, the process has reached a global or local minimum. The latter situation needs a strategy to recover, and we rely on repeating the beam search process using the best-known result as seed. This process continues until there is no improvement. The critical parameter here is the size $b$ of the beam and the implicit neighborhood. The use of beam search as an alternative to other schemes produces significant improvements in both recall and search speed, as supported experimentally in \S\ref{section/results}.

Please remember that the literature contains two ways to define the neighborhood; that is, the fixed neighborhood presented in NSW and the HNSW's SAT-based neighborhood over a large list of candidates. In this manuscript, an additional two variants are shown. The first strategy fixes the size of the neighborhood to $\left\lceil\log_2{n}\right\rceil$; where $n$ is the current number of indexed items, that is, the size of the neighborhood increases slowly as $n$ grows. This heuristic supposes that the neighbor is a logarithmic function of the size of the dataset. 
Furthermore, we also add a SAT-based neighborhood selection (see \S\ref{section/intro}) on the logarithmic-size list of neighbors, that is, the resulting neighborhood will be significantly smaller. We use these strategies to remove the tuning effort of selecting neighborhoods.

\subsubsection{Differences with previous approaches}
As explained, our BS uses the beam search strategy, starting with a random sample of the vertices that are improved iteratively. The process is repeated until the repetition does not improve the result set. The construction of the graph uses the search algorithm, as explained before; however, we introduce two heuristics Log and LogSat, that do not require an explicit parameter. In contrast, NSW requires to tune the number of greedy searches that are applied to solve the search, and also, the size of the neighborhood. On the other hand, HNSW requires to know the size of the neighborhood and the size of an internal list of candidates. Note that the HNSW also needs to maintain its hierarchy of layers; thus, our BS index has a more straightforward implementation. In summary, our BS-Log and BS-LogSat only need to set the size of the beam; as will be shown experimentally, both produce competitive performances for almost any of our benchmarks using a broad range of beam sizes.

It is necessary to remark that some properties like space complexity can be easily computed from parameters. The search-graph's memory is bounded by the neighborhood size and the size of the dataset; the explicit storage of the database is also a significant memory requirement. Log and LogSat have a variable-sized neighborhood, Log is logarithmic, and LogSat is upper bounded by Log. The search speed can be approximated in NSW since this property almost depends on the parameters, except for the number of hops that depends on the convergence. In our BeamSearch method, Algorithm
~\ref{alg/bs}, the search speed relies heavily on the convergence of the result, due to the beam search 
(BS) metaheuristic and the automatic repetition of the BS, lines 12-29. The search stops when the result's radius converges (line 15 for BS and line 29 for the repetition of BS); therefore, the dataset and the query are also a fundamental part of the search speed and the recall of each query.

Finally, our implementation of NSW is called IHC, and it accepts neighborhood strategies developed for BS.

\section{Experimental results}
\label{section/results}

In this section, we compare our techniques with the state of the art alternatives.
Our experiments were performed on an Intel(R) Xeon(R) CPU E5-2640 v3 @ 2.60GHz workstation with 16-core and 128 GiB of RAM, running Linux CentOS 7. We do not use the multiprocessing capabilities in the search process, and both indexes and databases are maintained in the main memory.
We select synthetic and real benchmarks. Unless otherwise specified, we report the performance of querying for $30$ nearest neighbors for all datasets. More detailed, the datasets and query sets are listed below:

\begin{itemize}
\item[--- \textsf{GIST-1M}.]
  This database contains one million 960-dimensional feature vectors \cite{jegou2011}.\footnote{The collection was retrieved from \url{http://corpus-texmex.irisa.fr/}} This collection obeys the computer vision modeling described in \cite{oliva2001modeling}, which were designed for scene recognition.
  We use the 1000 official queries for this benchmark; the Euclidean distance, $L_2$, is used for measuring the distance between point pairs. The average sequential search requires $0.385$ seconds on our testing machine. 

\item[--- \textsf{SIFT-1M} and \textsf{SIFT-100M}]
  We also used two more datasets from \cite{jegou2011}. These datasets are subsets of the one-billion dataset of Scale-invariant feature transform (SIFT) descriptors. Each descriptor consists of a 128-dimensional vector. We use the 10,000 official queries, solved with the Euclidean distance, i.e., $L_2$. An exhaustive search needs $2.540$ and $0.024$ seconds, in average, for \textsf{SIFT-100M} and \textsf{SIFT-1M}, respectively.

\item[--- \textsf{DEEPIMAGE-10M}.]
  The fourth real dataset is a ten-million subset of the one-billion dataset from a deep-learning-based image feature extraction~\cite{babenko2016efficient}. Each object is a 96-dimensional vector, and the distance notion is measured with the angle between vectors. The query collection has 10,000 vectors from the official query set; on average, each query is solved in $0.414$ seconds using a brute force solution.

\item[--- \textsf{RAND}.] We also generate several synthetic datasets in a unitary cube; these are standard benchmarks to study the algorithms' performance for a fixed size and dimension.
 i) Four datasets of dimension $8, 16, 32$, and $64$; each dataset contain three million vectors. A query is solved by exhaustive search in $0.035, 0.042, 0.069$, and $0.077$ seconds, respectively.
 ii) Two more datasets of 8-dimensional vectors of $300,000$ and $1,000,000$ elements; the exhaustive evaluation needs $0.005$ and $0.011$ seconds, respectively.
These datasets use $1000$ vectors for benchmarking, and the $L_2$ distance for measuring similarity.
\end{itemize}

We compare our indexes based on Beam Search (BS) and Iterated Hill Climbing (IHC) with HNSW, PQ, KNR, and LSH. For HNSW, PQ and LSH we used the implementation of FAISS,\footnote{Available at \url{https://github.com/facebookresearch/faiss}} written in C++; in particular, we use the cpu implementation. We also use FALCONN-LSH for benchmarking DEEPIMAGE.\footnote{Available at \url{https://github.com/FALCONN-LIB/FALCONN}.}
We use our own implementation for BS, IHC and KNR, written in the Julia language\footnote{\url{https://julialang.org/}} and it is also open-source.\footnote{Our source code is available at \url{https://github.com/sadit/SimilaritySearch.jl}}
We also show the performance Seq, our brute force search and  Flat, the FAISS's exhaustive search implementation.

We produce comparisons with several configurations for each index. These configurations must be adjusted for each dataset to reach optimal performance; however, we use several generic configurations that work for a wide range of problems. For instance, Beam Search (BS) uses three different beam sizes ($8, 16,$ and $32$) and three different strategies to compute the neighborhood: fixed neighborhood (FN) with 8, 16, and 32 neighbors; logarithmic (Log), and logarithmic with SAT (LogSat) neighborhoods. The same neighborhood strategies apply for IHC; it also uses three different restarting points ($8, 16, 32$ restarts). We test K nearest references (KNR) with $K=3,$, and $K=7$; in all cases, we fix the number of available references to be $\sqrt{n}$. We also show the performance of Product Quantization (PQ) with an Inverted Index, using several code sizes (using $8, 12,$, and $16$ encoding bits). LSH uses hash codes from 1024 and 2048 bits; we tested codes from 8 to 2048 bits, but the quality of the result set is degraded significantly for small codes.

Note that we compute all experiments using a single core; while allowing multi-threaded construction in large datasets (DEEPIMAGE-10M and SIFT-100M). Our multi-threading construction is almost straightforward and available in our open-source library; on the other hand, FAISS has native support for multi-threading construction. All our results are presented as the macro-average of our measurements; that is, we report the average of the result per each query. Please recall that our search and construction algorithms are entangled for BS and IHC; therefore, the construction time is proportional to the searching time and the size of the dataset.

\subsection{The effect of the dimension in the performance}
\label{sec/dimension}

\begin{figure}[!ht]
    \centering
    \includegraphics[width=\textwidth]{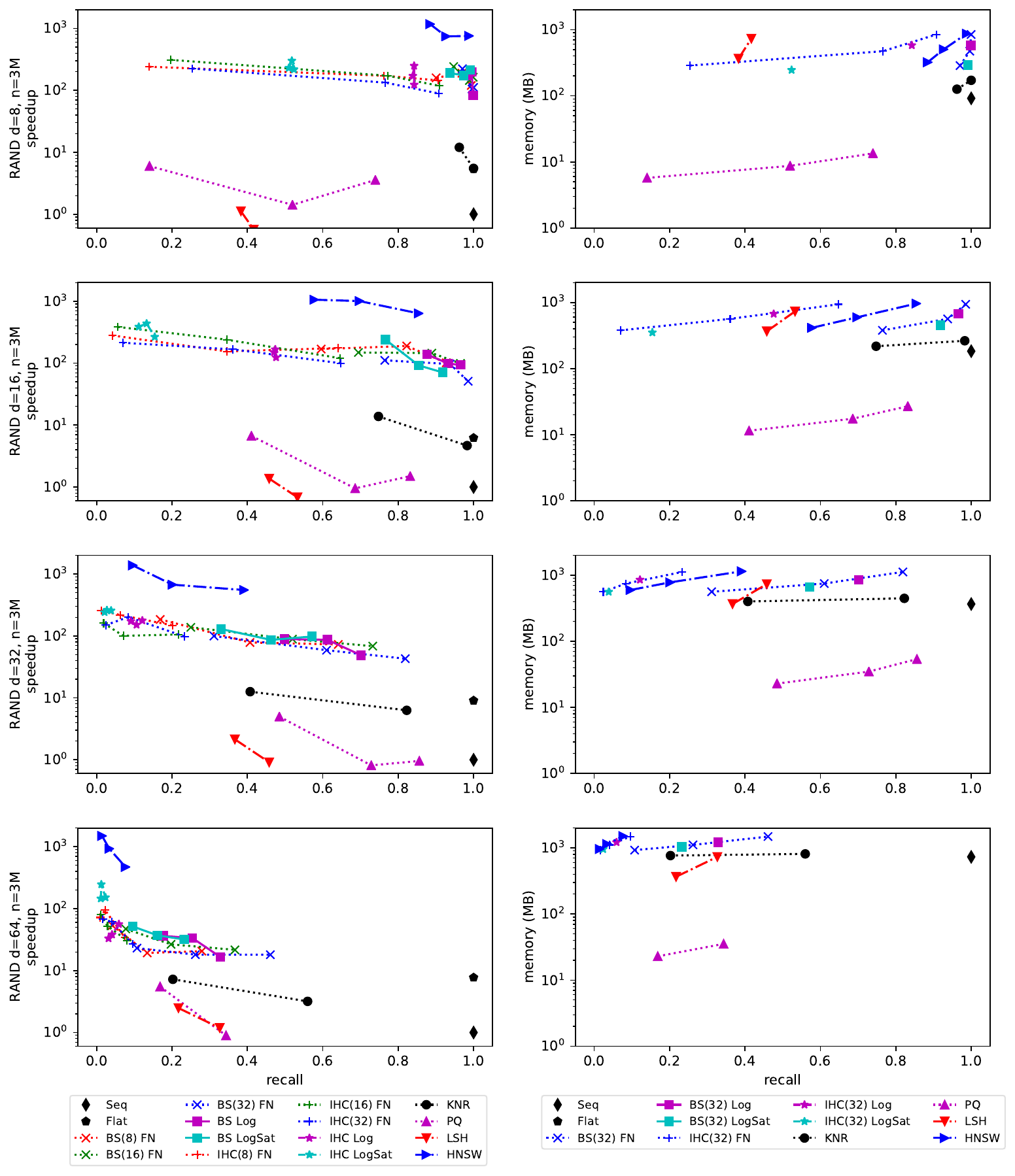}
    \caption{Performance comparison on synthetic datasets, 8, 16, 32, and 64 dimensions. On the left, recall vs. speedup (higher is better), on the right side, recall vs. memory (lower is better). Each row corresponds to a dataset, and the x-axis (recall) is shared between figure pairs.}
    \label{fig/rand-3M}
\end{figure}

Figure~\ref{fig/rand-3M} shows the performance of our synthetic 3 million benchmarks on different dimensions. The first row shows the performance of the indexes on RAND-3M (8-dimensional vectors); on the left side, we can observe the {\em speedup} in terms of our brute-force search implementation (logarithmic scale). The Flat index is 5.4 times faster than our exhaustive search; this speedup is related to SIMD implementations, and other low-level optimizations found in the FAISS library.
On the right side, we found the memory usage related to the recall of the first column. 

It is interesting to note how each index modifies its performance depending on the dimension of the dataset.
In the case of PQ, it improves the recall performance from 8 to 16, and from 16 to 32 dimensions. Since the speedup is reduced, the reason for the better recall can be explained as PQ being transformed on a kind of sequential scan as dimension increases; notoriously, the memory usage remains pretty low. KNR is an index that achieves relatively good recall scores with small memory footprints; however, its speedup is lower than graph-based indexes.

The first graph-based index in this comparison is the HNSW, which achieves faster searches; however, its recall is reduced as the dimension grows. Nonetheless, the memory usage of HNSW is among the largest ones. Our BS and IHC perform pretty differently depending on the neighborhood strategy. The fixed neighborhood (FN) for BS allows to achieve high recall values for eight and 16-dimensional vectors and remains high performant for the rest of our benchmarks. However, the memory footprints are the highest among the compared techniques. IHC-FN also achieves high recall values, yet never improves that of BS-FN. The logarithmic neighborhood (Log) also produces high recall values for almost all tested dimensions. This performance can suggest that the logarithmic strategy can be used as a rule of thumb in the selection of the neighborhood. Finally, the LogSat neighborhood strategy involves computing a Spatial Access Tree (SAT) grouping on the logarithmic neighborhood; LogSat achieves high recall scores using less memory than the Log neighborhood.

In general, it can be observed that the performance is degraded rapidly as dimension increases; however, the more resilient indexes are BS indexes and KNR indexes.

\subsection{Scalability}
\label{sec/scalability}

\begin{figure}[!ht]
    \centering
    \includegraphics[width=\textwidth]{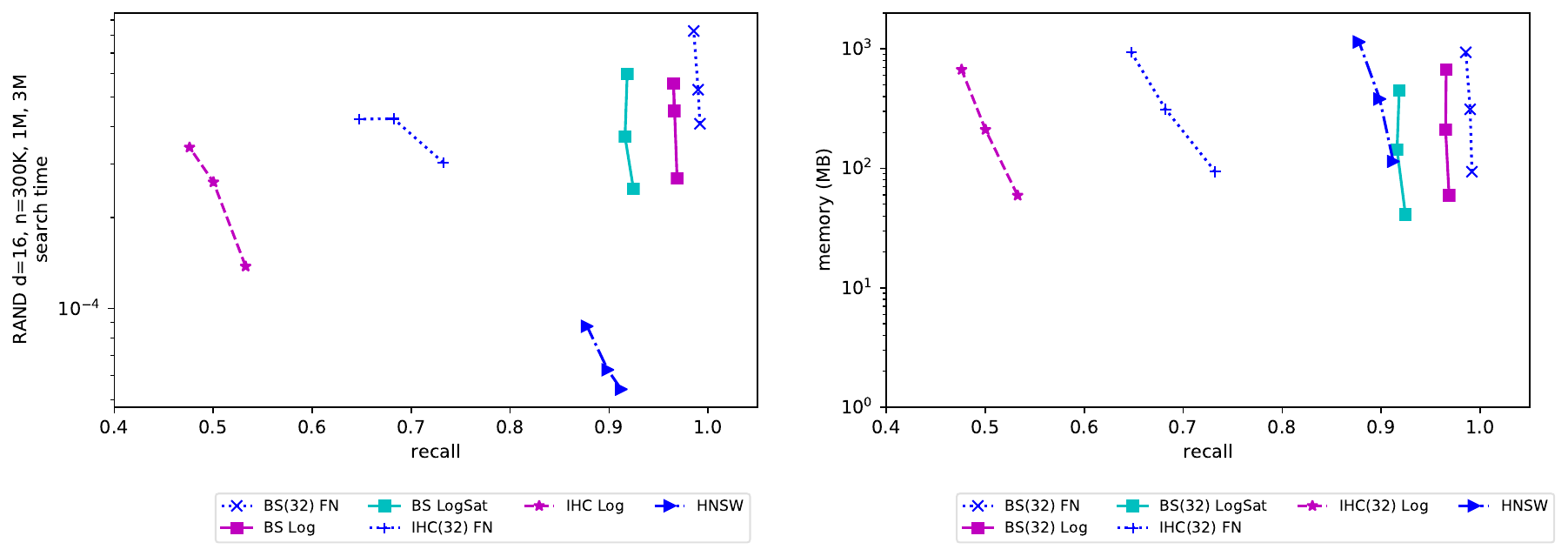}
    \caption{Performance comparison on our 16-dimensional synthetic dataset. On the left, recall vs. search time (higher is better), on the right side, recall vs. memory (lower is better). Each point corresponds to a different size on the database (300k, 1M, and 3M).}
    \label{fig/rand-16-var-size}
\end{figure}

In Figure~\ref{fig/rand-16-var-size}, we show the evolution of the search time of the indexes when the size of the dataset varies. Here, we used 16 dimensional random vectors with sizes of 300,000, 1,000,000 and 3,000,000 elements. For this experiment, we focus our comparison on BS, IHC, and HNSW with the idea of close ranges on the figures. On the left, we show the search time per query in seconds, on the right, the memory used. Each point on the series corresponds to a different size of the dataset. For each method, we selected the best configuration based on the recall. The search time graph shows that the indexes with better recall are the BS, but the HNSW is the fastest. An interesting property of the BS is that the recall is kept essentially constant despite the increase of the dataset size. Also, on the right, note how the needed memory for BS is less than that of HNSW.
In the rest of the section, we analyze the performance of real-world datasets.

\subsection{Performance on real datasets}
\begin{figure}[!ht]
    \centering
    \includegraphics[width=\textwidth]{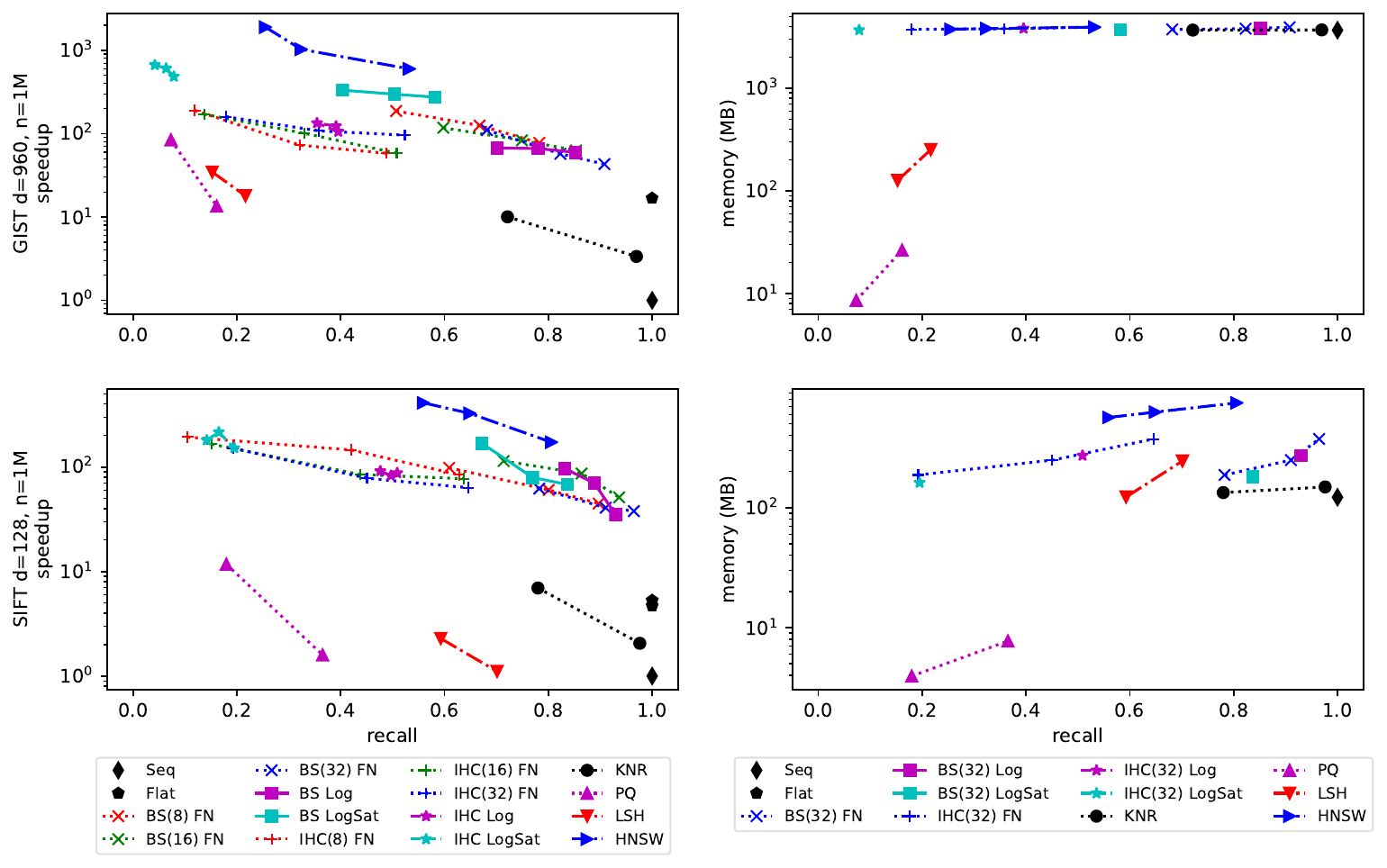}
    \caption{Performance comparison of our indexes with state-of-the-art alternatives in datasets with one million vectors. On the left side, recall vs. speedup (higher is better); on the right side, recall vs. memory (lower is better).}
    \label{fig/real-small}
\end{figure}

Figure~\ref{fig/real-small} shows the performance in two real datasets, GIST, and SIFT.
Each GIST vector contains one million 960-dimensional vectors, while SIFT contains one million 128-dimensional vectors.

The first row at Figure~\ref{fig/real-small} shows the performance for the GIST dataset; here, HNSW is the faster solution, yet it produces lower recall scores as compared with our graph-based alternatives. BS indexes achieve the highest ratings, and in particular, the faster among them is BS-LogSat. BS Log improves this recall with a small cost regarding speed. Note that both Log and LogSat are fixed strategies, such that a potential user only needs to set up the beam size, in the case of BS, and the number of restarts, in the case of IHC. These parameters can be tuned at the search stage; in contrast, fixed neighborhood strategies need more parameter tuning at the construction stage. We observe a speedup of 16.8 times from just using Flat instead of our exhaustive search. PQ and LSH obtain a competitive speed but very low recall. KNR is on the other extreme of the performance, i.e., high recall with low speeds.
Regarding memory usage, the dataset size (one million 960-dimensional vectors) dominates the requirements; this effect flattens memory curves, i.e., first-row of Figure~\ref{fig/real-small}.

The bottom-row of Figure~\ref{fig/real-small} compares the performance on the SIFT-1M dataset. Here we can observe a similar performance than before, HNSW is the faster index, but BS achieves better recall scores. In particular, BS-LogSat achieves a competitive performance by trading speed, recall, and memory conveniently; BS Log is a bit more precise but also a bit slower and has higher memory consumption.  The Flat search of FAISS is 5.3 times faster than our exhaustive search Seq; this speedup is similar to that found between BS and HSNW. Note the considerable difference between memory consumption among FAISS indexes and our implementations; this is because the highly optimized distance functions of FAISS require vectors of 32-bit floating-point numbers while we use vectors of 8-bit unsigned integers. These subtle details have a noticeable impact on the search performance; however, our gain is using four times lesser memory, and this saving will enable us to work with more massive datasets.

\subsection{Performance on large real-world datasets}
\label{sec/large}
\begin{figure}[!ht]
    \centering
    \includegraphics[width=\textwidth]{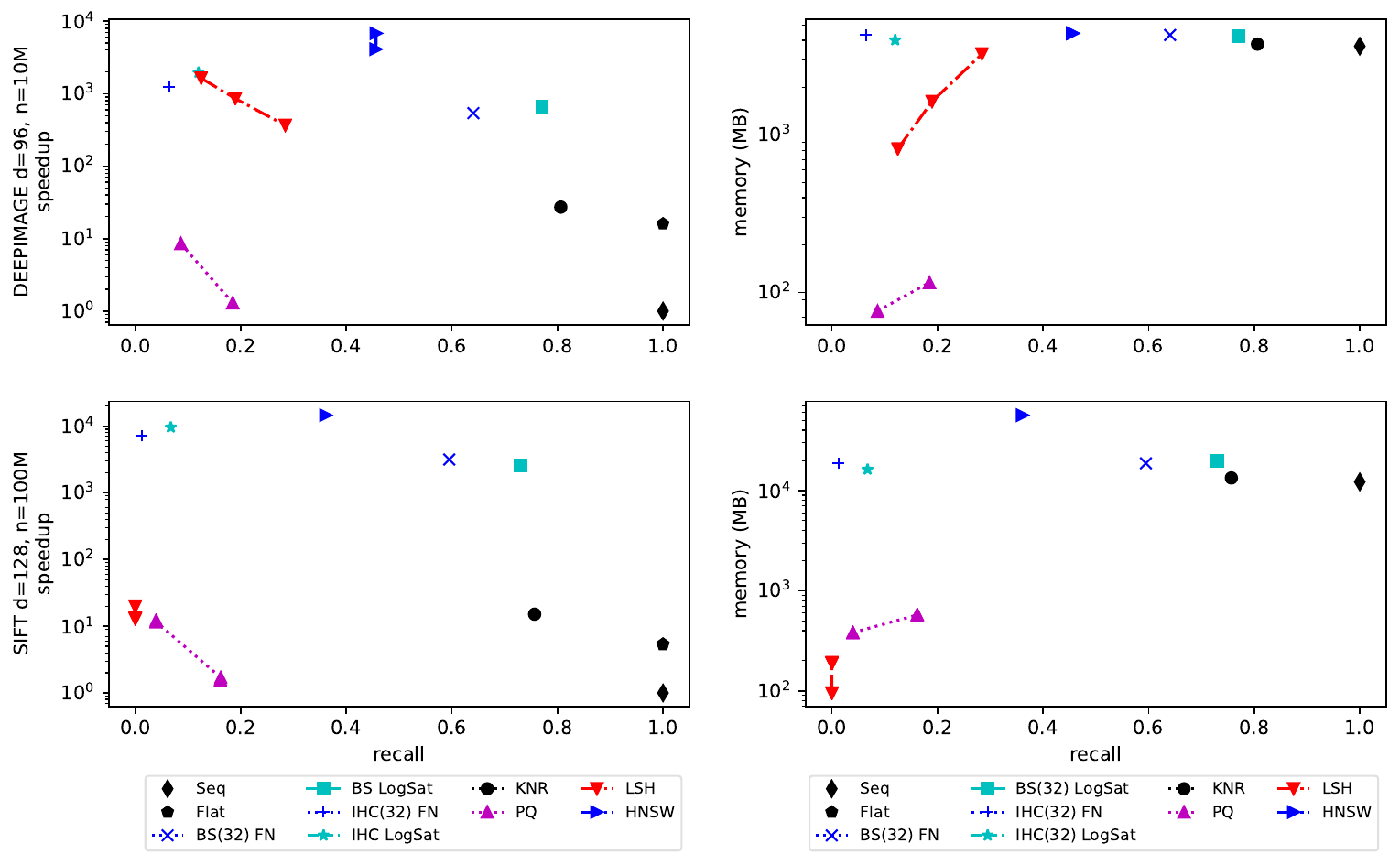}
    \caption{Performance comparison on two datasets: the DEEPIMAGE with ten, and SIFT, with one hundred million vectors. On the left side, recall vs. speedup (higher is better); on the right side, recall vs. memory (lower is better).}
    \label{fig/real-large}
\end{figure}

Large datasets require restricted memory indexes, which is particularly necessary for indexes that need to access the dataset to work. Our contributions fall into this index category, while others like LSH and PQ perform data projections that can alleviate the memory issues at the cost of recall or search speed reductions.

Figure~\ref{fig/real-large} compares the performance of search methods on two large datasets. In this set of experiments, we limit the memory usage of the indexes to use them.
We fix the memory size of BS and IHC indexes using Log, LogSat strategies that use a logarithmic neighborhood and a spatial-aware sample of a logarithmic neighborhood. We also compare a fixed neighborhood size of eight.
In particular, we restrict our comparison to BS with beam sizes of 32 and IHC with 32 random restarts. KNR uses three near references, and HNSW fixes a neighborhood of size eight.

In the first row, DEEPIMAGE, we can observe that memory usage is dominated by the database; and the speed dominated by HNSW. As in previous experiments, our BS-LogSat achieves competitive results in both recall and speed. The BS and IHC indexes use neighborhoods of size 8, with a beam size of 32 and the same number of restarts for IHC. Notoriously, KNR also exhibits higher recall scores with a small speed impact. Notably, the Flat index produces 10.5 times faster searches than our sequential search implementation. LSH and PQ are the indexes with relatively low memory consumption. In this benchmark, we use the LSH by \cite{andoni2015practical} since the distance function is the angle between vectors; we use 8, 16, and 32 tables and codes of 24 bits.

The search performance on the SIFT dataset with 100 million vectors is depicted in the bottom-row of Figure~\ref{fig/real-large}. As before, HNSW is the faster index, yet it has higher memory requirements due to the use of 32-bit floating-point numbers. On the other hand, our BS-LogSat achieves the best recall while maintaining a competitive search speed. PQ and LSH have a pretty small memory-footprint but at the cost of both speed and result quality.

LSH and PQ indexes have a small memory footprint; others like HNSW solve queries quickly and accurately. Our BS indexes achieve competitive tradeoffs between speed, memory, and recall. This equilibrium is particularly suitable for computationally limited environments or huge workloads. 

\subsection{On the construction cost}
The construction time is a wall on many metric indexes. For instance, using 32 threads under our workstation 32 with 16 cores and hyper-threading. We enabled multithreading for our indexes, FAISS implementation also supports multithreading construction natively. All indexes were set to use 32 threads. 
The construction of DEEPIMAGE-10M needs 397.4 seconds for BS-LogSat, 265.8 seconds for IHC-LogSat, 152.0 seconds for HNSW, and 60 seconds for KNR. Both PQ and LSH use a few seconds to create the full index. The construction of SIFT-100M requires 228 minutes for BS-LogSat, 105 minutes for IHC-LogSat, 96 minutes for HNSW, 33 minutes for PQ, and less than a minute for LSH.
In general, we can observe that faster and more accurate indexes require more construction resources and time. Please note that the speed differences between HSNW and BS are not preserved on the construction time, even when both are based on similar principles. In other words, the simpler  structure behind BS-LogSat reduces the gap with the HNSW in the construction time as compared with the gap in the search cost.

\subsection{The performance impact of query size}
\begin{table}[t]

\caption{BS's recall and speedup performances for RAND-3M and several number of nearest neighbors $k$. In both cases, higher values are better; i.e., recall goes from 0 to 1, and speedup is a positive score. Best scores per dataset and $k$ are marked in bold.
\label{tab/var-k}}

\resizebox{\textwidth}{!}{
\begin{tabular}{llrr rrrr rrrr}
\toprule
       &    & \multicolumn{5}{c}{recall} & \multicolumn{5}{c}{speedup} \\
       \cmidrule(r){3-7} \cmidrule(l){8-12}
method & $\mathcal{N}$ & $k=1$ & $k=3$ & $k=10$ & $k=30$ & $k=100$ & $k=1$ & $k=3$ & $k=10$ & $k=30$ & $k=100$\\ \midrule
\multicolumn{12}{c}{$d=8, n=3 \times 10^6$} \\ 
\midrule
BS(8) & Log & 0.92 & 0.98 & 0.99 & 0.98 & 0.96 & 121.8 & 123.9 & 126.0 & 84.5 & 51.5\\
BS(16) & Log &\bftab 0.92 & 0.98 & 1.00 & 1.00 & 0.99 & 80.0 & 89.4 & 108.9 & 73.1 & 42.4\\
BS(32) & Log & 0.91 &\bftab 0.99 &\bftab 1.00 &\bftab 1.00 &\bftab 1.00 & 113.2 & 66.6 & 64.6 & 62.4 & 41.6\vspace{1mm}\\
BS(8) & LogSat & 0.71 & 0.92 & 0.95 & 0.94 & 0.89 & 200.9 & \bftab 211.9 & 94.8 & 86.4 & 51.3\\
BS(16) & LogSat & 0.71 & 0.92 & 0.98 & 0.97 & 0.94 & 204.0 & 141.8 & \bftab 140.3 &\bftab 124.4 &\bftab 69.1\\
BS(32) & LogSat & 0.71 & 0.93 & 0.98 & 0.99 & 0.98 &\bftab 209.5 & 113.8 & 94.6 & 77.1 & 43.2\\
\midrule
\multicolumn{12}{c}{$d=16, n=3 \times 10^6$} \\ 
\midrule
BS(8) & Log & 0.54 & 0.79 & 0.87 & 0.88 & 0.87 & 156.5 & 63.5 & 99.3 & 73.2 & 42.0\\
BS(16) & Log & 0.52 & 0.81 & 0.92 & 0.93 & 0.92 & 126.8 & 68.9 & 74.6 & 61.1 & 35.0\\
BS(32) & Log &\bftab 0.56 &\bftab 0.82 &\bftab 0.95 &\bftab 0.97 &\bftab 0.95 & 146.8 & 63.4 & 34.2 & 39.7 & 27.3\vspace{1mm}\\
BS(8) & LogSat & 0.31 & 0.61 & 0.74 & 0.77 & 0.77 & 188.3 &\bftab 165.6 & \bftab 139.2 &\bftab 89.1 &\bftab 52.5\\
BS(16) & LogSat & 0.37 & 0.65 & 0.82 & 0.85 & 0.84 & 132.1 & 62.8 & 56.5 & 49.3 & 34.2\\
BS(32) & LogSat & 0.37 & 0.67 & 0.87 & 0.92 & 0.91 &\bftab 215.8 & 104.3 & 59.7 & 57.0 & 38.6\\
\midrule
\multicolumn{12}{c}{$d=32, n=3 \times 10^6$} \\ 
\midrule
BS(8) & Log & 0.16 & 0.29 & 0.43 & 0.50 & 0.56 & 98.1 & 80.6 & 87.4 & 65.6 & 47.2\\
BS(16) & Log & 0.14 & 0.35 & 0.52 & 0.61 & 0.66 & 60.0 & 34.6 & 48.1 & 49.0 & 32.2\\
BS(32) & Log &\bftab 0.16 &\bftab 0.36 &\bftab 0.59 &\bftab 0.70 &\bftab 0.74 & 130.4 & 42.2 & 40.0 & 41.4 & 32.1\vspace{1mm}\\
BS(8) & LogSat & 0.05 & 0.16 & 0.26 & 0.33 & 0.41 & 83.7 & 93.4 &\bftab 97.1 & 80.9 & 46.1\\
BS(16) & LogSat & 0.06 & 0.19 & 0.37 & 0.46 & 0.53 &\bftab 276.6 &\bftab 103.5 & 95.8 &\bftab 83.7 &\bftab 59.1\\
BS(32) & LogSat & 0.09 & 0.23 & 0.44 & 0.58 & 0.63 & 213.2 & 97.8 & 57.3 & 63.9 & 41.1\\
\midrule
\multicolumn{12}{c}{$d=64, n=3 \times 10^6$} \\ 
\midrule
BS(8) & Log & 0.04 & 0.09 & 0.13 & 0.18 & 0.25 & 36.0 & 41.6 & 35.5 & 22.2 & 12.7\\
BS(16) & Log & 0.04 & 0.11 & 0.19 & 0.25 & 0.33 & 37.0 & 28.5 & 26.0 & 18.5 & 11.8\\
BS(32) & Log &\bftab 0.05 &\bftab 0.12 &\bftab 0.23 &\bftab 0.33 &\bftab 0.41 & 90.1 & 34.3 & 19.4 & 17.1 & 10.5\vspace{1mm}\\
BS(8) & LogSat & 0.02 & 0.04 & 0.07 & 0.10 & 0.15 &\bftab 192.8 &\bftab 133.2 &\bftab 103.3 &\bftab 59.8 &\bftab 28.9\\
BS(16) & LogSat & 0.02 & 0.06 & 0.11 & 0.16 & 0.23 & 161.1 & 81.8 & 63.3 & 42.7 & 24.9\\
BS(32) & LogSat & 0.02 & 0.06 & 0.15 & 0.23 & 0.31 & 177.6 & 85.5 & 36.6 & 29.2 & 16.7\\
\bottomrule
\end{tabular}
}

\end{table}

This experiment focuses on describing the performance of BS and the number of neighbors to be retrieved, $k$; we focus on BS and two neighborhood strategies Log and LogSat.
Due to the conditional statement of line 22 of Alg.~\ref{alg/bs}, the number of neighbors that we search controls the convergence of the process. That is, small $k$ values will converge faster than large values since small $k$ values will less likely to improve after some steps. While this implies that for small $k$ values, it may be better to search for a larger number of neighbors and then cut to $k$, it is useful to know the dynamic of this parameter on the search performance.

Table~\ref{tab/var-k} shows the recall and speedup performances of our BS method while searching for a different number of neighbors $k$. The table lists the performance for several beam sizes on RAND-3M, i.e., with dimensions $8,16,32,$ and $64$.

For almost all benchmarks, the best recall scores are achieved with Log neighborhood, and in particular, for $b=32$. Nonetheless, this high-performance configuration will have a relatively higher memory cost than the LogSat strategy, as described in our experimental comparison. On the other hand, LogSat improves Log strategy regarding speedup; this is also observed in previous experiments. The small neighborhood produced by LogSat is the cause of this speedup improvement. It is also noticeable that small $b$ values will produce fast searches, and large $b$ values will produce higher recall scores.

Regarding $k$, Table~\ref{tab/var-k} shows a minimal effect for small dimensions, but it becomes an issue as the dimension grows. For instance, for BS-Log with $b=32$ and $k=1$ we achieve a recall of $0.92$, when $d=16$, the recall reduces to $0.56$, and for $d=32$ it reduces significantly to $0.16$; for dimension $64$, it becomes as low as $0.05$. The impact for queries with higher $k$ is lesser, for instance, when $k=100$ the recall score goes from $1.00$ ($d=8$) to $0.41$ ($d=64$), for BS(32)-Log.

\begin{table}[t]
\centering
\caption{Recall scores for RAND-3M datasets for $k=1$ queries with expanded $k'$; high recall values are better. Best scores per $k'$ and dimension are marked in bold.
\label{tab/expand-k}}
\scalebox{0.8}{
\begin{tabular}{ll rrrr}
\toprule
method    & $\mathcal{N}$ & $d=8$ & $d=16$ & $d=32$ & $d=64$ \\
\midrule
\multicolumn{6}{c}{$k'=3$} \\
\midrule
BS(8)  & Log    &\bftab 0.99 & 0.83 & 0.32 & 0.10 \\
BS(16) & Log    &\bftab 0.99 & 0.83 & 0.38 & 0.12 \\
BS(32) & Log    &\bftab 0.99 &\bftab 0.86 &\bftab 0.40 &\bftab 0.13\vspace{1mm}\\
BS(8)  & LogSat & 0.95 & 0.63 & 0.17 & 0.05 \\
BS(16) & LogSat & 0.94 & 0.69 & 0.21 & 0.06 \\
BS(32) & LogSat & 0.95 & 0.70 & 0.28 & 0.06 \\
\midrule
\multicolumn{6}{c}{$k'=10$} \\
\midrule
BS(8)  & Log    &\bftab 1.00 & 0.92 & 0.55 & 0.18 \\
BS(16) & Log    &\bftab 1.00 & 0.97 & 0.66 & 0.26 \\
BS(32) & Log    &\bftab 1.00 &\bftab 0.99 &\bftab 0.69 &\bftab 0.30\vspace{1mm}\\
BS(8)  & LogSat & 0.99 & 0.84 & 0.32 & 0.08 \\
BS(16) & LogSat &\bftab 1.00 & 0.92 & 0.45 & 0.14 \\
BS(32) & LogSat &\bftab 1.00 & 0.94 & 0.56 & 0.20 \\
\midrule
\multicolumn{6}{c}{$k'=30$} \\
\midrule
BS(8)  & Log    &\bftab 1.00 & 0.97 & 0.72 & 0.16 \\
BS(16) & Log    &\bftab 1.00 & 0.99 & 0.81 & 0.23 \\
BS(32) & Log    &\bftab 1.00 &\bftab 1.00 &\bftab 0.89 & 0.35\vspace{1mm}\\
BS(8)  & LogSat &\bftab 1.00 & 0.93 & 0.47 & 0.42 \\
BS(16) & LogSat &\bftab 1.00 & 0.97 & 0.66 & 0.53 \\
BS(32) & LogSat &\bftab 1.00 & 0.99 & 0.78 &\bftab 0.63 \\
\midrule
\multicolumn{6}{c}{$k'=100$} \\
\midrule
BS(8)  & Log    &\bftab 1.00 & 0.99 & 0.84 & 0.42 \\ 
BS(16) & Log    &\bftab 1.00 &\bftab 1.00 & 0.93 & 0.53 \\ 
BS(32) & Log    &\bftab 1.00 &\bftab 1.00 &\bftab 0.95 &\bftab 0.63\vspace{1mm}\\
BS(8)  & LogSat &\bftab 1.00 & 0.97 & 0.66 & 0.27 \\ 
BS(16) & LogSat &\bftab 1.00 & 0.99 & 0.82 & 0.36 \\ 
BS(32) & LogSat &\bftab 1.00 &\bftab 1.00 & 0.90 & 0.51 \\ 
\bottomrule
\end{tabular}
}
\end{table}

As commented before, it is possible to palliate the effect whenever queries consist of few neighbors, i.e., small $k$, using a larger $k'$ to search and then cut the fetched result to $k$. Table~\ref{tab/expand-k} illustrates how this strategy improves $k=1$ (nearest neighbor queries) for different values of $k'$ and several synthetic datasets.
For instance, for 8-dimensional vectors, expanding to $k'=3$ produces a small but noticeable improvement as compared to $k'=1$ (first column in Table~\ref{tab/var-k}) for Log neighborhood and a more significant improvement for LogSat strategy. A close to perfect recall is achieved with $k'=10$, and this is remarkable for BS-LogSat since it barely surpasses 0.7 recall scores without query expansion. For $d=16$, the recall scores were also improved for both BS-Log and BS-LogSat, with respect to $k'=1$; here, we also obtain better recalls for larger $k'$, more noticeable for BS-LogSat.
Both 32 and 64-dimensional datasets show lower scores even when $k' \geq 10$; however, in both dimensions, using $k' > 1$ produces significant improvements over $k'=1$, see Table~\ref{tab/var-k}. Nonetheless, the later performances are caused by the difficulty of searching in high dimensional datasets, even with this relatively enlarged $k'$.

\begin{table}[t]
\centering
\caption{Recall performance for $k=30$ with expanded query $k'=100$ for four different dimensional datasets, i.e., RAND-3M. Best recall scores per dataset are marked in bold.
\label{tab/expand-k-30-100}}
\scalebox{0.8}{
\begin{tabular}{ll rrrr}
\toprule
method  &$\mathcal{N}$ & $d=8$ & $d=16$ & $d=32$ & $d=64$ \\
\midrule
BS(8)  &Log    &\bftab 1.00 & 0.95 & 0.66 & 0.29 \\
BS(16) &Log    &\bftab 1.00 & 0.97 & 0.77 & 0.39 \\
BS(32) &Log    &\bftab 1.00 &\bftab 0.99 &\bftab 0.84 &\bftab 0.48\vspace{1mm}\\
BS(8)  &LogSat &       0.98 & 0.88 & 0.50 & 0.18 \\
BS(16) &LogSat &       0.99 & 0.94 & 0.62 & 0.27 \\
BS(32) &LogSat &\bftab 1.00 & 0.97 & 0.73 & 0.36 \\ \bottomrule
\end{tabular}
}
\end{table}

Please note that increasing $k$ to alleviate the convergence problems works mainly for whenever $k$ is small. Table~\ref{tab/expand-k-30-100} illustrates this effect for $k=30$ and $k'=100$. For $d=8$ and $d=16$, the recall scores remain practically unchanged. We observed a small improvement, for $d=32$ and $d=64$ and specially for BS-LogSat. Based on this, the expansion of $k'$ seems to have a small impact for relatively large $k$ queries.

\begin{table}[ht!]
\centering
\caption{Recall scores for BS(32) indexes for different queries without query expansion for SIFT-1M, GIST-1M, and DEEPIMAGE-10M datasets.
\label{tab/real-dataset-k}}
\resizebox{1.0\textwidth}{!}{
\large
\begin{tabular}{l rrrrr rrrrr rrrrr}
\toprule
              & \multicolumn{15}{c}{recall} \\ \cmidrule{2-16}
$\mathcal{N}$ & \multicolumn{5}{c}{SIFT $d=128, n=10^6$} & \multicolumn{5}{c}{GIST $d=960, n=10^6$} & \multicolumn{5}{c}{DEEPIMAGE $d=96, n=10^7$} \\
              \cmidrule(r){2-6} \cmidrule(lr){7-11} \cmidrule(l){12-16}
              & $k\!=\!1$ & $k\!=\!3$  & $k\!=\!10$ & $k\!=\!30$  & $k\!=\!100$
              & $k\!=\!1$ & $k\!=\!3$  & $k\!=\!10$ & $k\!=\!30$  & $k\!=\!100$
              & $k\!=\!1$ & $k\!=\!3$  & $k\!=\!10$ & $k\!=\!30$  & $k\!=\!100$
              \\
\cmidrule(r){2-6} \cmidrule(lr){7-11} \cmidrule(l){12-16} 

Log    & 0.51 & 0.76 & 0.90 & 0.93 & 0.93 &   0.39 & 0.65 & 0.81 & 0.85 & 0.85 &   & & & & \\
LogSat & 0.26 & 0.55 & 0.76 & 0.84 & 0.84 &   0.09 & 0.29 & 0.47 & 0.58 & 0.62 &   0.22 & 0.51 & 0.71 & 0.78 & 0.79 \\
\bottomrule
\end{tabular}
}
\end{table}

Table~\ref{tab/real-dataset-k} shows the recall performance on varying $k$ of BS(32), one of our best configurations, for real datasets. We can observe that nearest neighbor queries ($k=1$) achieve pretty low recall scores for SIFT and GIST with LogSat neighborhood.
Please recall that BS-Log uses a logarithmic neighborhood, and BS-LogSat takes a small sample among the logarithmic neighborhood using a spatial-aware covering method; therefore, BS-LogSat always produce smaller or equal neighborhoods than Log, and using it implies a tradeoff among quality and memory.
We can observe, in all datasets, how as $k$ increases the recall may also increase; this performance is linked to the convergence of the search algorithm, as explained before.
For DEEPIMAGE-10M we did not present results for Log neighborhood strategy due to our self-imposed memory limits for large datasets, see \S\ref{sec/large}.

\begin{table}[ht!]
\centering
\caption{Recall scores for BS(32) indexes for $k=1$ queries using query expansion $k'$ for SIFT-1M, GIST-1M, and DEEPIMAGE-10M datasets. 
\label{tab/real-dataset-var-kprime}}

\resizebox{1.0\textwidth}{!}{
\begin{tabular}{lrrrr rrrr rrrr}
\toprule
              & \multicolumn{12}{c}{recall} \\ \cmidrule{2-13}
$\mathcal{N}$
& \multicolumn{4}{c}{SIFT $d=128, n=10^6$}
& \multicolumn{4}{c}{GIST $d=960, n=10^6$}
& \multicolumn{4}{c}{DEEPIMAGE $d=96, n=10^7$} \\
              \cmidrule(r){2-5} \cmidrule(lr){6-9} \cmidrule(l){10-13}
              & $k'\!=\!3$ & $k'\!=\!10$ & $k'\!=\!30$ & $k'\!=\!100$
              & $k'\!=\!3$ & $k'\!=\!10$ & $k'\!=\!30$ & $k'\!=\!100$
              & $k'\!=\!3$ & $k'\!=\!10$ & $k'\!=\!30$ & $k'\!=\!100$
            \\
            \cmidrule(r){2-5} \cmidrule(lr){6-9} \cmidrule(l){10-13}
Log    & 0.79 & 0.94 & 0.98 & 0.99 & 0.67 & 0.86 & 0.93 & 0.96 & - & - & - & - \\
LogSat & 0.58 & 0.83 & 0.94 & 0.97 & 0.30 & 0.50 & 0.70 & 0.81 & 0.53 & 0.78 & 0.89 & 0.93 \\
\bottomrule
\end{tabular}
}
\end{table}

As commented before, it is hard to achieve high recall for $k=1$ since $k$ is essential for convergence in Alg.~\ref{alg/bs}. Here, the query expansion technique can improve the result quality. Table~\ref{tab/real-dataset-var-kprime} shows the recall performance for $k=1$ queries and expansion $k'$ of $3, 10, 30, $ and $100$.
Note that $k=1$ queries improve its recall scores as compared to not expand queries; for instance, using $k'=100$ has a remarkable impact for SIFT-1M, it goes from a score of 0.51 to 0.99 for BS(32)-Log; and from 0.26 to 0.97 using BS(32)-LogSat. The recall for GIST-1M ($k=1$) is improved consistently as $k'$ increases, and the effect for BS(32)-LogSat is noticeable since improves from 0.09 to 0.81; the Log neighborhood achieves 0.96 when $k'=100$. 
The DEEPIMAGE-10M also significantly improves its expected performance using query expansion, going from a recall of 0.22 to 0.89 for $k'=30$. Notoriously, $k'=100$ produces a recall of $0.93$.

\section{Conclusions}
\label{sec/conclusions}

In this manuscript, we adapted the Beam Search (BS) metaheuristic to solve nearest neighbor queries and introduce the use of logarithmic neighborhoods (BS-Log) to create the underlying search graph. Moreover, we also introduce an additional method BS-LogSat based on selecting a small subset of the logarithmic neighborhood. We give a broad review of the related state of the art and provide an insightful explanation of our methods. 
Our primary effort was to improve the search speed without reducing the quality of the search results and producing relatively small data structures. These objectives were reached for several datasets, all of them with a plethora of dimensions, sizes, and domains. We also aimed to reduce the number of parameters for the indexes, since parameter tuning puts a significant barrier to potential users; for instance, our BS-Log and BS-LogSat need only a single parameter to work, i.e., the size of the beam. An efficient and effective index with minimal user intervention is elusive. Still, we believe our work is a milestone in this direction, and this goal deserves effort from the research community.

Our experimental study found that our indexes achieve competitive performances in most of our benchmarking datasets, as compared with state-of-the-art alternatives. The memory requirements of our graph are  smaller than alternative search algorithms reaching similar performances. 

Similarly, it is interesting to notice that HNSW is blazing fast; this is due to its high-performance algorithm and its highly optimized implementations using SIMD instructions, a set of instructions found in modern CPUs.
Our indexes are also open-source and written in the Julia language. Distance functions are optimized through the Julia native machinery, that in many cases, achieves competitive implementations similar to that produced by many low-level optimizations. In our experiments, we observed a sustained five to ten-fold speedup improvement in the brute force implementation of FAISS and ours. Despite this, our BS indexes are also competitive in many aspects. Our implementation also accepts user-defined distance functions smoothly and efficiently; this feature can be exploited in exploratory data analysis and other data-science tasks.

Since our indexes are constructed incrementally, the insertion operation is native; however, the deletion algorithm is not yet studied. These features are useful for applications using metric databases to represent knowledge, like those using incremental learning. The literature in metric indexes used to solve the deletion operation marking those items as unavailable for most operations \cite{CNBYMacmcs2001}. 
In other approaches, the metric indexes can afford real deletions. However, to maintain the invariants of the indexes, a partial or even a complete reconstruction of the index is needed (\cite{DSATjea08,DSATsisap09,M-trees}).
A recent approach uses the strategy of Bentley and Saxe (\cite{NaidanMagnusIS2014}) to produce dynamic structures utilizing a list of $\log n$ static structures.
The similarity search through the combinatorial optimization approach is robust enough to support deletions, straightforwardly. However, the approach requires more research on determining optimal replacements of the removed items, and if any other operations are needed to ensure high-quality results after a large number of deletions. This operation deserves attention in future research.

\section*{Acknowledgements}
The authors would like to thank the anonymous reviewers for their valuable comments and suggestions to improve the quality of this manuscript. 

\bibliographystyle{alpha}
\bibliography{super-bib}
\end{document}